\documentclass[11pt]{article}
\usepackage{harvard,float,latexsym}
\usepackage{epsfig,psfrag}
\usepackage{amsmath,amssymb,amsthm}
\newtheorem*{lemma}{Lemma}
\def\th{{\rm th}}
\def\E{{\rm E}}
\def\P{{\rm P}}

\makeatletter  
\renewcommand{\verbatim@font}{%
  \ttfamily\small\catcode`\<=\active\catcode`\>=\active%
}
\makeatother   

\begin{document}

\title{Fast Computation of the Median by Successive Binning}
\author{Ryan J. Tibshirani\\
Dept. of Statistics, Stanford University, Stanford, CA 94305\\ 
email: {\tt ryantibs@stanford.edu}}
\date{October 14, 2008}
\maketitle

\begin{abstract}
In many important problems, one uses the median instead of the mean to estimate a population's center, since the former is more robust.  But in general, computing the median is considerably slower than the standard mean calculation, and a fast median algorithm is of interest.  The fastest existing algorithm is \textsf{quickselect}.  We investigate a novel algorithm \textsf{binmedian}, which has $O(n)$ average complexity.  The algorithm uses a recursive binning scheme and relies on the fact that the median and mean are always at most one standard deviation apart.  We also propose a related median approximation algorithm \textsf{binapprox}, which has $O(n)$ worst-case complexity.  These algorithms are highly competitive with \textsf{quickselect} when computing the median of a single data set, but are significantly faster in updating the median when more data is added.
\end{abstract}

\section{Introduction}\label{sec:intro}
In many applications it is useful to use the median instead of the mean to measure the center of a population.  Such applications include, but are in no way limited to: biological sciences; computational finance; image processing; and optimization problems.  Essentially, one uses the median because it is not sensitive to outliers, but this robustness comes at a price: computing the median takes much longer than computing the mean.  Sometimes one finds the median as a step in a larger iterative process (like in many optimization algorithms), and this step is the bottleneck.  Still, in other problems (like those in biological applications), one finds the median of a data set and then data is added or subtracted.  In order to find the new median, one must recompute the median in the usual way, because the standard median algorithm cannot utilize previous work to shortcut this computation.  This brute-force method leaves much to be desired.  

We propose a new median algorithm \textsf{binmedian}, which uses buckets or ``bins'' to recursively winnow the set of the data points that could possibly be the median.  Because these bins preserve some information about the original data set, the \textsf{binmedian} algorithm can use previous computational work to effectively update the median, given more (or less) data.  For situations in which an exact median is not required, \textsf{binmedian} gives rise to an approximate median algorithm \textsf{binapprox}, which rapidly computes the median to within a small margin of error.  We compare \textsf{binmedian} and \textsf{binapprox} to \textsf{quickselect}, the fastest existing algorithm for computing the median.  We first describe each algorithm in detail, beginning with \textsf{quickselect}.

\section{Quickselect}\label{sec:quickselect}
\subsection{The Quickselect Algorithm}
\textsf{Quickselect} \cite{quickselect} is not just used to compute the median, but is a more general algorithm to select the $k^\th$ smallest element out of an array of $n$ elements.  (When $n$ is odd the median is the $k^\th$ smallest with $k=(n+1)/2$, and when $n$ is even it is the mean of the elements $k=n/2$ and $k=n/2+1$.)  The algorithm is derived from \textsf{quicksort} \cite{quicksort}, the well-known sorting algorithm.  First it chooses a ``pivot'' element $p$ from the array, and rearranges (``partitions'') the array so that all elements less than $p$ are to its left, and all elements greater than $p$ are to its right.  (The elements equal to $p$ go on either side).  Then it recurses on the left or right subarray, depending on where the median element lies.  In pseudo-code:

\vspace{0.2in}
\noindent
Quickselect$(A,k)$:
\begin{enumerate}
\item Given an array $A$, choose a pivot element $p$
\item Partition $A$ around $p$; let $A_1,A_2,A_3$ be the subarrays of points $<,=,> p$
\item If $k \leq \mathrm{len}(A_1)$: return Quickselect$(A_1,k)$
\item Else if $k > \mathrm{len}(A_1) + \mathrm{len}(A_2)$: return Quickselect$\left(A_3, k - \mathrm{len}(A_1) - \mathrm{len}(A_2)\right)$
\item Else: return $p$
\end{enumerate}

The fastest implementations of \textsf{quickselect} perform the steps 3 and 4 iteratively (not recursively), and use a standard in-place partitioning method for step 2.  On the other hand, there isn't a universally accepted strategy for choosing the pivot $p$ in step 1.  For the comparisons in sections \ref{sec:comparisons-1} and \ref{sec:comparisons-2}, we use a very fast Fortran implementation of \text{quickselect} taken from \citeasnoun{numerical-recipes-Fortran}.  It chooses the pivot $p$ to be the median of the bottom, middle, and top elements in the array.  

\subsection{Strengths and Weaknesses of Quickselect}
There are many advantages to \textsf{quickselect}.  Not only does it run very fast in practice, but when the pivot is chosen randomly from the array, \textsf{quickselect} has $O(n)$ average computational complexity \cite{quickselectAC}.  Another desirable property is that uses only $O(1)$ space, meaning that the amount of extra memory it needs doesn't depend on the input size.  Perhaps the most underappreciated strength of \textsf{quickselect} is its simplicity.  The algorithm's strategy is quite easy to understand, and the best implementations are only about 30 lines long.

One disadvantage of \textsf{quickselect} is that it rearranges the array for its own computational convenience, necessitated by the use of in-place partitioning.  This seemingly innocuous side-effect can be a problem when maintaining the array order is important (for example, when one wants to find the median of a column of a matrix).  To resolve this problem, \textsf{quickselect} must make a copy of this array, and use this scratch copy to do its computations (in which case it uses $O(n)$ space).   

But the biggest weakness of \textsf{quickselect} is that it is not well-suited to problems where the median needs to be updated with the addition of more data.  Though it is able to efficiently compute the median from an in-memory array, its strategy is not conducive to saving intermediate results.  Therefore, if data is appended to this array, we have no choice with \textsf{quickselect} but to compute the median in the usual way, and hence repeat much of our previous work.  This will be our focus in section \ref{sec:comparisons-2}, where we will give an extension of \textsf{binmedian} that is able to deal with this update problem effectively.

\section{Binmedian}\label{sec:binmedian}
\subsection{The Binmedian Algorithm}
First we give a useful lemma:
\begin{lemma}\label{lemma}
If $X$ is a random variable having mean $\mu$, variance $\sigma^2$, and median $m$, then $m \in \lbrack \mu-\sigma, \mu+\sigma \rbrack$.
\end{lemma}
\begin{proof} Consider
\begin{align*}
|\mu - m| &= |\E(X - m)| \\
&\leq \E|X-m| \\
&\leq \E|X-\mu| \\
&= \E\sqrt{(X-\mu)^2} \\
&\leq \sqrt{\E (X-\mu)^2} \\
&= \sigma.
\end{align*}
Here the first inequality is due to Jensen's inequality, and the second inequality is due to the fact that the median minimizes the function $\Phi(a) = \E|X-a|$.  The third inequality is due to the concave version of Jensen's inequality.
\end{proof}

Given data points $x_1,\ldots x_n$ and assuming that $n$ is odd\footnote{An analogous strategy holds when $n$ is even.  Now at each stage we have to keep track of the bin that contains the ``left'' median (the $k^\th$ smallest element with $k=n/2$) and the bin that contains the ``right'' median ($k=n/2+1$).  This is more tedious than the odd case, but is conceptually the same, so hereafter we'll assume that $n$ is odd whenever we talk about \textsf{binmedian}.}, the basic strategy of the $\textsf{binmedian}$ algorithm is:
\begin{enumerate}
\item Compute the mean $\mu$ and standard deviation $\sigma$
\item Form $B$ bins across $\lbrack \mu-\sigma, \mu+\sigma \rbrack$, map each data point to a bin
\item Find the bin $b$ that contains the median
\item Recurse on the set of points mapped to $b$
\end{enumerate}
Here $B$ is a fixed constant, and in step 2 we consider $B$ equally spaced bins across $\lbrack \mu-\sigma, \mu+\sigma \rbrack$.  That is, we consider intervals of the form $$\mu-\sigma + \left\lbrack \frac{i}{B} \cdot 2\sigma, ~ \frac{i+1}{B} \cdot 2\sigma \right)$$ for $i=0,1,\ldots B-1$.  Iterating over the data $x_1,\ldots x_n$, we count how many points lie in each one of these bins, and how many points lie to the left of these bins.  We denote these counts by $N_i$ and $N_L$.  

In step 3, to find the median bin we simply start adding up the counts from the left until the total is $\geq (n+1)/2$.  That is, $b$ is minimal such that $$N_L + \sum_{i=0}^b N_i \geq \frac{n+1}{2}.$$  Note that our lemma guarantees that $b \in \{0,\ldots B-1\}$.    
             
Finally, in step 4 we iterate through $x_1,\ldots x_n$ once again to determine which points were mapped to bin $b$.  Then we consider $B$ equally spaced bins across $\mu-\sigma + \left\lbrack \frac{b}{B} \cdot 2\sigma, ~ \frac{b+1}{B} \cdot 2\sigma \right)$, and continue as before (except that now $N_L$ is the number of points to the left of the median bin, i.e. $N_L \leftarrow N_L+\sum_{i=0}^{b-1} N_i$).  We stop when there is only one point left in the median bin -- this point is the median.  (In practice, it is actually faster to stop when the remaining number of points in the median bin is $\leq C$, another fixed constant, and then find the median directly using insertion sort.)

Notice that a larger value of $B$ gives fewer points that map to the median bin, but it also means more work to find $b$ at each recursive step.  For the time trials in sections \ref{sec:comparisons-1} and \ref{sec:comparisons-2}, we choose $B=1000$ because it weighs these two factors nicely.  We also replace the recursive step with an iterative loop, and use $C=20$.  \textsf{Binmedian} is implemented in Fortran for the time trials, and this code, as well as an equivalent implementation in C, is freely available at http://stat.stanford.edu/\verb1~1ryantibs/median/.

\subsection{Strengths and Weaknesses of Binmedian}
Beginning with its weaknesses, the \textsf{binmedian} algorithm will run faster with fewer points in the median's bin, and therefore its runtime is necessarily dependent on the distribution of the input data.  The worst case for \textsf{binmedian} is when there is a huge cluster of points surrounding the median, with a very large standard deviation.  With enough outliers, the median's bin could be swamped with points for many iterations.

However, we can show that the algorithm's asymptotic running time is largely unaffected by this problem: with very reasonable assumptions on the data's distribution, \textsf{binmedian} has $O(n)$ average complexity (see the Appendix).  Moreover, this robustness also seems to be true in practice.  As we shall see in section \ref{sec:comparisons-1}, the algorithm's runtime is only moderately affected by unfavorable distributions, and in fact, \textsf{binmedian} is highly competitive with \textsf{quickselect} in terms of running time.  

Like \textsf{quickselect}, the \textsf{binmedian} algorithm uses $O(1)$ space, but (like \textsf{quickselect}) this requires it to change the order of the input data.  The greatest strength of \textsf{binmedian} is that the algorithm is able to quickly cache some information about the input data, when it maps the data points to bins.  This gives rise to a fast median approximation algorithm, \textsf{binapprox}.  It also enables \textsf{binmedian}, as well as \textsf{binapprox}, to efficiently recompute the median when we are given more data.  We'll see this in section \ref{sec:comparisons-2}, but first we discuss \textsf{binapprox} in the next section.   

\section{Binapprox}\label{sec:binapprox}
\subsection{The Binapprox Algorithm}
In some situations, we do not need to compute the median exactly, and care more about getting a quick approximation.  \textsf{Binapprox} is a simple approximation algorithm derived from \textsf{binmedian}.  It follows the same steps as \textsf{binmedian}, except that it stops once it has computed the median bin $b$, and just returns the midpoint of this interval.  Hence to be perfectly clear, the algorithm's steps are:
\begin{enumerate}
\item Compute the mean $\mu$ and standard deviation $\sigma$
\item Form $B$ bins across $\lbrack \mu-\sigma, \mu+\sigma \rbrack$, map each data point to a bin
\item Find the bin $b$ that contains the median
\item Return the midpoint of bin $b$
\end{enumerate}

The median can differ from this midpoint by at most $1/2$ the width of the interval, or $1/2 \cdot 2\sigma/B = \sigma/B$.  Since we use $B=1000$ in practice, \textsf{binapprox} is accurate to within $1/1000^\th$ of a standard deviation.

Again, we implement \textsf{binapprox} in Fortran for sections \ref{sec:comparisons-1} and \ref{sec:comparisons-2}.  This code as well as C code for \textsf{binapprox} is available at http://stat.stanford.edu/\verb1~1ryantibs/median/.

\subsection{Strengths and Weaknesses of Binapprox}
Unlike \textsf{binmedian}, the runtime of \textsf{binapprox} doesn't depend on the data's distribution, since it doesn't perform the recursive step.  It requires $O(1)$ space, and doesn't perturb the input data, so rearranging is never a problem.  The algorithm has $O(n)$ worst-case computational complexity, as it only needs 3 passes through the data.  It consistently runs faster than \textsf{quickselect} and \textsf{binmedian} in practice.  Most importantly it extends to a very fast algorithm to deal with the update problem.
 
\textsf{Binapprox}'s main weakness is fairly obvious: if the standard deviation is extremely large, the reported approximation could be significantly different from the actual median.

\section{Speed Comparisons on Single Data Sets}\label{sec:comparisons-1}
We compare \textsf{quickselect}, \textsf{binmedian}, and \textsf{binapprox} across data sets coming from various distributions.  For each data set we perform a sort and take the middle element, and report these runtimes as a reference point for the slowest way to compute the median.  

We test a total of eight distributions.   The first four --- uniform over $[0,1]$, normal with $\mu=0,\sigma=1$, exponential with $\lambda=1$, chi-square with $k=5$ --- are included because they are fundamental, and frequently occur in practice.  The last four are mixed distributions that are purposely unfavorable for the \textsf{binmedian} algorithm.  They have a lot of points around the median, and a huge standard deviation.  Two of these are mixtures of standard normal data and uniformly distributed data over $[-10^3,10^3]$, resp. $[-10^4,10^4]$.  The other two are mixtures of standard normal data and exponential data with $\lambda=10^{-3}$, resp. $\lambda=10^{-4}$.  All these mixtures are even, meaning that an equal number of points come from each of the two distributions.  Table \ref{tab:times-1} gives the results of all these timings, which were performed on an Intel Xeon 64 bit, 3.00GHz processor.   
 
\begin{table}[htbp]
\centering
\begin{tabular}{|l||r|r||r|r|}
\hline
& \multicolumn{2}{c||}{\boldmath $U(0,1)$} & \multicolumn{2}{c|}{\boldmath $N(0,1)$} \\
\hline
& time & ratio & time & ratio \\
\hline 
\textsf{Quickselect} & 107.05 (0.47) & 1.17 & 99.70 (1.05) & 1.17 \\
\hline
\textsf{Binmedian} & 112.45 (0.55) & 1.23 & 106.95 (1.51) & 1.25 \\
\hline
\textsf{Binapprox} & 91.30 (0.32) & 1 & 85.35 (0.86) & 1 \\
\hline
\textsf{Sort} & 3205.84 (5.97) & 35.11 & 3231.22 (4.24) & 37.86 \\
\hline
\hline
& \multicolumn{2}{c||}{\boldmath $E(1)$} & \multicolumn{2}{c|}{\boldmath $\chi^2_5$} \\
\hline 
& time & ratio & time & ratio \\
\hline 
\textsf{Quickselect} & 106.75 (0.83) & 1.62 & 107.80 (4.29) & 1.32 \\
\hline
\textsf{Binmedian} & 87.25 (0.96) & 1.32 & 104.65 (4.12) & 1.28 \\
\hline
\textsf{Binapprox} & 65.90 (0.75) & 1 & 81.75 (3.97) & 1 \\
\hline
\textsf{Sort} & 3217.51 (5.40) & 48.82 & 3225.82 (4.70) & 39.46 \\
\hline
\hline
& \multicolumn{2}{c||}{\bf \boldmath $N(0,1)$ and $U(-10^3,10^3)$} & \multicolumn{2}{c|}{\bf \boldmath $N(0,1)$ and $U(-10^4,10^4)$} \\
\hline
& time & ratio & time & ratio \\
\hline 
\textsf{Quickselect} & 103.65 (1.15) & 1.33 & 101.00 (0.96) & 1.30  \\
\hline
\textsf{Binmedian} & 126.65 (1.20) & 1.62 & 130.90 (1.25) & 1.69 \\
\hline
\textsf{Binapprox} & 78.15 (1.02) & 1 & 77.50 (1.28) & 1 \\
\hline
\textsf{Sort} & 3225.40 (5.58) & 41.27 & 3214.16 (3.89) & 41.47 \\
\hline
\hline
& \multicolumn{2}{c||}{\bf \boldmath $N(0,1)$ and $E(10^{-3})$} & \multicolumn{2}{c|}{\bf \boldmath $N(0,1)$ and $E(10^{-4})$} \\
\hline
& time & ratio & time & ratio \\
\hline 
\textsf{Quickselect} & 98.45 (1.13) & 1.59 & 100.90 (1.23) & 1.58 \\
\hline
\textsf{Binmedian} & 88.90 (1.63) & 1.43 & 106.85 (0.89) & 1.67 \\
\hline
\textsf{Binapprox} & 62.10 (1.16) & 1 & 63.80 (1.25) & 1 \\
\hline
\textsf{Sort} & 3228.93 (5.56) & 52.00 & 3236.71 (2.95) & 50.73 \\
\hline
\end{tabular}
\caption[tab:times-1]{\em Runtimes in milliseconds of \textsf{quickselect}, \textsf{binmedian}, \textsf{binapprox}, and sort for computing the median of $1+10^7$ data points from different distributions.  We performed 10 repetitions for each distribution, and report the average and standard deviation of the runtimes.  (In fact, for each repetition we drew 20 different data sets from the distribution, and timed how long it takes for the algorithm to sequentially compute the medians of these 20 data sets, in one contiguous block.  Then we divided this number by 20; this was done to avoid timing inaccuracies.)  We also list the runtime ratios, relative to the fastest time.} 
\label{tab:times-1}
\end{table} 
 
For the first four distributions, \textsf{quickselect} and \textsf{binmedian} are very competitive in terms of runtime, with \textsf{quickselect} doing better on the data from $U(0,1)$ and $N(0,1)$, and \textsf{binmedian} doing better on the data from $E(1)$ and $\chi^2_5$.  It makes sense that \textsf{binmedian} does well on the data drawn from $E(1)$ and $\chi^2_5$, upon examining the probability density functions of these distributions.  The density function of the exponential distribution, for example, is at its highest at 0 and declines quickly, remaining in sharp decline around the median ($\log{2}/\lambda$).  This implies that there will be (relatively) few points in the median's bin, so \textsf{binmedian} gets off to a good start.

Generally speaking, \textsf{binmedian} runs slower than \textsf{quickselect} over the last four distributions.  But considering that these distributions are intended to mimic \textsf{binmedian}'s degenerate case (and are extreme examples, at that), the results are not too dramatic at all.  The \textsf{binmedian} algorithm does worse with the normal/uniform mixtures (skewed on both sides) than it does with the normal/exponential mixtures (skewed on one side).  At its worst, \textsf{binmedian} is 1.3 times slower than \textsf{quickselect} (130.9 versus 101 ms, on the $N(0,1)$ and $U(-10^4,10^4)$ mixture), but is actually slightly {\em faster} than \textsf{quickselect} on the $N(0,1)$ and $E(10^{-3})$ mixture.

\textsf{Binapprox} is the fastest algorithm across every distribution.  It maintains a reasonable margin on \textsf{quickselect} and \textsf{binmedian} (up to 1.62 and 1.69 times faster, respectively), though the differences are not striking.  In the next section, we focus on the median update problem, where \textsf{binmedian} and \textsf{binapprox} display a definite advantage. 

\section{Speed Comparisons on the Update Problem}\label{sec:comparisons-2}
Suppose that we compute the median of some data $x_1,\ldots x_{n_0}$, and then we're given more data $x_{n_0+1},\ldots x_n$, and we're asked for the median of the aggregate data set $x_1,\ldots x_n$.  Of course we could just go ahead and compute the median of $x_1,\ldots x_n$ directly, but then we would be redoing much of our previous work to find the median of $x_1,\ldots x_{n_0}$.  We'd like a better strategy for updating the median.   

Consider the following: we use \textsf{binmedian} to compute the median of $x_1,\ldots x_{n_0}$, and save the bin counts $N_i$, $i=0,\ldots B-1$, and $N_L$, from the first iteration.  We also need save the mean $\mu_0$ and the standard deviation $\sigma_0$.  Given $x_{n_0+1},\ldots x_n$, we map these points to the original $B$ bins and just increment the appropriate counts.  Then we compute the median bin in the usual way: start adding $N_L + N_0 + N_1 + \ldots$ and stop when this is $\geq (n+1)/2$.   

But now, since we haven't mapped the data $x_1,\ldots x_n$ to bins over its proper mean and standard deviation, we aren't guaranteed that the median lies in one of the bins.  In the case that the median lies to the left of our bins (i.e. $N_L \geq (n+1)/2$) or to the right of our bins (i.e. $N_L + N_0 + \ldots N_{B-1} < (n+1)/2$), we have to start all over again and perform the usual \textsf{binmedian} algorithm on $x_1,\ldots x_n$.  But if the median does lie in one of the bins, then we can continue onto the second iteration as usual.  This can provide a dramatic savings in time, because we didn't even have to touch $x_1,\ldots x_{n_0}$ in the first iteration.

We can use \textsf{binapprox} in a similar way: save $\mu_0,\sigma_0,N_i,N_L$ from the median computation on $x_1,\ldots x_{n_0}$, and then use them to map $x_{n_0+1},\ldots x_n$.  We then determine where the median lies: if it lies outside of the bins, we perform the usual \textsf{binapprox} on $x_1,\ldots x_n$.  Otherwise we can just return midpoint of the median's bin.  

In this same situation, \textsf{quickselect} does not have a way to save its previous work, and must redo the whole computation.\footnote{One might argue that \textsf{quickselect} sorts half of the array $x_1,\ldots x_{n_0}$, and it could use this sorted order when given $x_{n_0+1},\ldots x_n$ to expedite the partitioning steps, and thus quickly compute the median of $x_1,\ldots x_n$.  First, keeping this sorted order does not actually provide much of a speed up.  Second, and most importantly, that \textsf{quickselect} might keep $x_1,\ldots x_{n_0}$ in the order that it produced is an unrealistic expectation in practice.  There may be several other steps that occur between initially computing the median and recomputing the median, and these steps could manipulate the order of the data.  Also, if the data's order shouldn't be perturbed in the first place (for example, if we're computing the median of a column of a matrix), then \textsf{quickselect} has to make a copy of $x_1,\ldots x_{n_0}$, and only rearranges the order of this copy.  It's not practical to keep this copy around until we're given $x_{n_0+1},\ldots x_n$, even if $n_0$ is only reasonably large.}  In table \ref{tab:times-2}, we demonstrate the effectiveness of \textsf{binmedian} and \textsf{binapprox} on this update problem.  Instead of adding data $x_{n_0+1},\ldots x_n$ only once, we add it many times, and we record how long it takes for the algorithms to compute these medians in one contiguous block.  \textsf{Binmedian} and \textsf{binapprox} each keep a global copy of $\mu_0,\sigma_0,N_i,N_L$.  They set these globals given the initial data $x_1,\ldots x_{n_0}$, and as more data is added, they dynamically change them if the median ever lies outside of the bins. 

\begin{table}[htbp]
\centering
\begin{tabular}{|l|r|r|}
\hline
\multicolumn{3}{|c|}{\bf \boldmath $1+10^7$ pts $\sim N(0,25)$ and } \\
\multicolumn{3}{|c|}{\bf \boldmath then $20 \times (10^5$ pts $\sim N(0,25))$} \\
\hline
& time & ratio \\
\hline 
\textsf{Quickselect} & 2236 (8.23) & 22.36 \\
\hline
\textsf{Binmedian} & 584 (5.27) & 5.84 \\
\hline
\textsf{Binapprox} & 100 (9.42) & 1 \\
\hline
\hline
\multicolumn{3}{|c|}{\bf \boldmath $1+10^7$ pts $\sim N(0,25)$ and } \\
\multicolumn{3}{|c|}{\bf \boldmath then $20 \times (10^5$ pts $\sim N(2,4))$} \\
\hline
& time & ratio \\
\hline 
\textsf{Quickselect} & 2234 (7.07) & 24.55 \\
\hline
\textsf{Binmedian} & 563 (10.75) & 6.19 \\
\hline
\textsf{Binapprox} & 91 (12.65) & 1 \\
\hline
\hline
\multicolumn{3}{|c|}{\bf \boldmath $1+10^6$ pts $\sim N(0,25)$ and } \\
\multicolumn{3}{|c|}{\bf \boldmath then $20 \times (10^6$ pts $\sim N(1/2\cdot j,25), j=1,\ldots 20)$} \\
\hline
& time & ratio \\
\hline 
\textsf{Quickselect} & 2429 (11.36) & 18.99 \\
\hline
\textsf{Binmedian} & 616 (9.94) & 4.81 \\
\hline
\textsf{Binapprox} & 128 (5.27) & 1 \\
\hline
\hline
\multicolumn{3}{|c|}{\bf \boldmath $1+10^6$ pts $\sim N(0,25)$ and } \\
\multicolumn{3}{|c|}{\bf \boldmath then $20 \times (10^6$ pts $\sim N(10,25))$} \\
\hline
& time & ratio \\
\hline 
\textsf{Quickselect} & 2376 (16.19) & 19.01 \\
\hline
\textsf{Binmedian} & 629 (11.01) & 5.03 \\
\hline
\textsf{Binapprox} & 125 (5.68) & 1 \\
\hline
\end{tabular}
\caption[tab:times-2]{\em Runtimes in milliseconds of \textsf{quickselect}, \textsf{binmedian}, and \textsf{binapprox} for the median update problem.  The algorithms initially compute the median of a data set, then we sequentially add more data, and the algorithms must recompute the median after each addition.  Whenever new data is added, \textsf{quickselect} computes the median of the aggregate data set with its usual strategy, and hence repeats much of its previous work.  On the other hand, \textsf{binmedian} and \textsf{binapprox} can use their cached bin counts to effectively update the median.  For each distribution, we performed 10 repetitions, and we list the average and standard deviation of these times.  We give the runtime ratios, which are taken relative to the fastest time.}
\label{tab:times-2}
\end{table}

The first two situations from table \ref{tab:times-2} represent the best case scenario for \textsf{binmedian} and \textsf{binapprox}.  The added data sets are fairly small when compared to the initial data set ($n-n_0=10^5$ data points versus $n_0=1+10^7$ data points).  Also, the added data lies mainly inside the initial bin range $[\mu_0-\sigma_0, \mu_0+\sigma_0]$, so the median will never lie outside of the bins.  This means that \textsf{binmedian} and \textsf{binapprox} never have to undo any of their previous work (i.e. they never have to recompute the mean or standard deviation of the data set, and re-bin all the points), and are able update the median very quickly.  At their best, \textsf{binmedian} is faster than \textsf{quickselect} by a factor of 3.97, and \textsf{binapprox} is faster by a factor of 24.55.   

The last two situations represent realistic, not-so-ideal cases for \textsf{binmedian} and \textsf{binapprox}.  The added data sets are equal in size to the initial data set ($n-n_0=10^6$ and $n_0=1+10^6$).  Moreover, the added data lies mainly outside of the initial bin range, which means that the median will eventually lie outside of the bins, forcing \textsf{binmedian} and \textsf{binapprox} to recompute the mean and standard deviation and re-bin all of the data points.  (This may happen more than once.)  Nonetheless, \textsf{binmedian} and \textsf{binapprox} still maintain a very good margin on \textsf{quickselect}, with \textsf{binmedian} running up to 3.94 times faster than \textsf{quickselect} and \textsf{binapprox} running up to 19.01 times faster.

It is important to note that when \textsf{binapprox} successfully updates the median (that is, it finds the new median bin without having to recompute the mean, standard deviation, etc.), it is accurate to within $\sigma_0/1000$, where $\sigma_0$ is the standard deviation of the original data set $x_1,\ldots x_{n_0}$.  To bound its accuracy in terms of the standard deviation $\sigma$ of the whole data set $x_1,\ldots x_n$, consider 
\begin{equation*}
\sigma = \sqrt{ \frac{1}{n} \sum_{i=1}^n (x_i - \mu)^2 } 
\geq \sqrt{ \frac{1}{n} \sum_{i=1}^{n_0} (x_i - \mu)^2 } 
\geq \sqrt{ \frac{1}{n} \sum_{i=1}^{n_0} (x_i - \mu_0)^2 }
= \sqrt{\frac{n_0}{n}} \sigma_0
\end{equation*}
where $\mu_0$ is the mean of $x_1,\ldots x_{n_0}$ and $\mu$ is the mean of $x_1,\ldots x_n$.  Therefore, \textsf{binapprox}'s error in this situation is at most \begin{equation}\label{var-inequality}
\frac{\sigma_0}{1000} \leq \frac{\sqrt{\frac{n}{n_0}}\sigma}{1000}.
\end{equation}
If $n = 2n_0$, this becomes $\sqrt{2}\sigma/1000 \approx \sigma/700$, which is still quite small. 
 
Finally, observe that we can use an analogous strategy for \textsf{binmedian} and \textsf{binapprox} to recompute the median after {\it removing} (instead of adding) a subset of data.  Suppose that, having computed the median of $x_1,\ldots x_n$, we wish to know median when $x_{n_0+1},\ldots x_n$ are removed from the data set.  Then as before, we save $\mu,\sigma,N_i,N_L$ from our median computation on $x_1,\ldots x_n$, map $x_{n_0+1},\ldots x_n$ to bins, and decrement (instead of increment) the counts appropriately. 
 
\section{Discussion and Summary}
We have proposed an algorithm \textsf{binmedian}, which computes the median by repeated binning on a selectively smaller subset of the data, and a corresponding approximation algorithm \textsf{binapprox}.  \textsf{Binmedian} has $O(n)$ average complexity when mild assumptions are made on the data's distribution function, and \textsf{binapprox} has $O(n)$ worst-case complexity.  For finding the median of a single data set, \textsf{binmedian} is highly competitive with \textsf{quickselect}, although it becomes slower when there are many points near the median and the data has an extremely large standard deviation.  On single data sets, \textsf{binapprox} is consistently faster no matter the distribution.  \textsf{Binmedian} and \textsf{binapprox} both outperform \textsf{quickselect} on the median update problem, wherein new data is successively added to a base data set; on this problem, \textsf{binapprox} can be nearly 25 times faster than \textsf{quickselect}.  

In most real applications, an error of $\sigma/1000$ (an upper bound for \textsf{binapprox}'s error) is perfectly acceptable.  In many biological applications this can be less than the error attributed to the machine that collects the data.  Also, we emphasize that \textsf{binapprox} performs all of its computations without side effects, and so it does not perturb the input data in any way.  Specifically, \textsf{binapprox} doesn't change the order of the input data, as do \textsf{quickselect} and \textsf{binmedian}.  For applications in which the input's order needs to be preserved, \textsf{quickselect} and \textsf{binmedian} must make a copy of the input array, and this will only increase the margin by which \textsf{binapprox} is faster (in tables \ref{tab:times-1} and \ref{tab:times-2}).  

Finally, it is important to note the strategy for \textsf{binmedian} and \textsf{binapprox} is well-suited to situations in which computations must be distributed.  The computations to find the mean and standard deviation can be easily distributed -- hence with the data set divided into many parts, we can compute bin counts separately on each part, and then share these counts to find the median bin.  For \textsf{binapprox} we need only to report the midpoint of this bin, and for \textsf{binmedian} we recurse and use this same distributed strategy.  This could work well in the context of wireless sensor networks, which have recently become very popular in many different areas of research, including various military and environmental applications (see \citeasnoun{wsn}).


For any situation in which an error of $\sigma/1000$ is acceptable, there seems to be no downside to using the \textsf{binapprox} algorithm.  It is in general a little bit faster than \textsf{quickselect} and it admits many nice properties, most notably its abilities to rapidly update the median and to be distributed.  We hope to see diverse applications that benefit from this.

\section*{Appendix}
\subsection*{Proof that Binmedian Has $O(n)$ Expected Running Time}
Let $X_1,\ldots X_n$ be iid points from a continuous distribution with density $f$.  Assuming that $f$ is bounded and that $\E(X_1^4)<\infty$, we show that the \textsf{binmedian} algorithm has expected $O(n)$ running time.\footnote{Notice that this expectation is taken over draws of the data $X_1,\ldots X_n$ from their distribution.  This is different from the usual notion of expected running time, which refers to an expectation taken over random decisions made by the algorithm, holding the data fixed.}

Now for notation: let $f \leq M$, and $\E(X_1^2) = \sigma^2$.  Let $\hat\mu$ and $\hat\sigma$ be the empirical mean and standard deviation of $X_1,\ldots X_n$.  Let $N_j$ be the number of points at the $j^{\th}$ iteration ($N_0=n$), and let $T$ be the number of iterations before termination.  We can assure $T \leq n$.\footnote{To do this we make a slight modification to the algorithm, which does not change the running time.  Right before the recursive step (step 4), we check if all of the data points lie in the median bin.  If this is true, then instead of constructing $B$ new bins across the bin's endpoints, we construct the bins across the minimum and maximum data points.  This assures that at least one data point will be excluded at each iteration, so that $T \leq n$.}

At the $j^{\th}$ iteration, the algorithm performs $N_j+B$ operations;
therefore its runtime is
$$R = \sum_{j=0}^T(N_j + B) \leq \sum_{j=0}^n (N_j + B) = \sum_{j=0}^n N_j + B(n+1)$$
and its expected runtime is
$$\E(R) \leq \sum_{j=0}^n \E(N_j) + B(n+1).$$

Now we compute $\E(N_j)$. Let $b$ denote the median bin; this is a
(random) interval of length $2\hat\sigma/B$.  Well 
\begin{align*}
\E(N_1) &= n \P(X_1 \in b) \\
&\leq n\P(X_1 \in b; ~\hat\sigma \leq \sigma + \epsilon) ~+ ~n\P(\hat\sigma > \sigma+\epsilon).
\end{align*}
Well $\P(X_1 \in b; ~\hat\sigma \leq \sigma + \epsilon) < M\cdot
2(\sigma+\epsilon)/B$ (see the picture), and $\P(\hat\sigma-\sigma > \epsilon) = C/n$ (see the lemma; this is why we need a finite fourth moment of $X_1$).  Therefore $$\E(N_1) \leq \frac{n}{B} 2M(\sigma+\epsilon) + C.$$

\begin{figure}[tb]
\centering
\begin{psfrags}
\psfrag{M}{$M$}
\psfrag{f}{$f$}
\psfrag{2s/B}{$\displaystyle \frac{2\hat\sigma}{B}$}
\psfrag{<=2(s+e)/B}{$\displaystyle \leq \frac{2(\sigma+\epsilon)}{B}$}
\epsfig{file=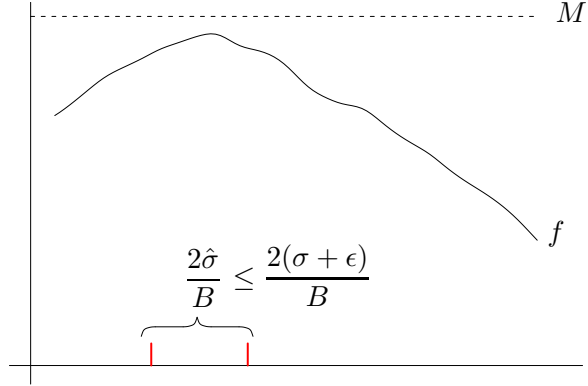,width=3in}
\end{psfrags}
\caption{The area under the curve between the two red lines is $\leq
  M\cdot 2(\sigma+\epsilon)/B$.}
\end{figure}

In general $$\E(N_j) = n\P(X_1 \in b, X_1 \in b_2, \ldots \mbox{ and }
X_1 \in b_j) = n\P(X_1 \in b_j)$$ where $b_i$ is the median's bin at the $i^\th$
iteration, so that we can similarly bound
$$\E(N_j) \leq \frac{n}{B^j} 2M(\sigma+\epsilon) + C.$$

Therefore
\begin{align*}
\E(R) &\leq \sum_{j=0}^n\left(\frac{n}{B_j}2M(\sigma+\epsilon)+C\right) + B(n+1) \\
     &= 2M(\sigma+\epsilon)\left(\sum_{j=0}^n 1/B^j\right)n + (B+C)(n+1) \\
     &\leq 2M(\sigma+\epsilon)\left(\sum_{j=0}^\infty 1/B^j\right)n + (B+C)(n+1) \\
     &= O(n).
\end{align*}

\vspace{0.2in}
\begin{lemma} With the same assumptions as above,
  $\P(\hat\sigma-\sigma>\epsilon) = O(n^{-1})$ for
  any $\epsilon>0$.
\end{lemma}
\begin{proof} Notice
\begin{align*}
\left\{\hat\sigma-\sigma>\epsilon\right\} &=
\left\{(\hat\sigma-\sigma)^2 > \epsilon^2 ; ~\hat\sigma>\sigma\right\}
\\
&= \left\{\hat\sigma^2 -2\hat\sigma\sigma +\sigma^2 > \epsilon^2 ;
  ~\hat\sigma>\sigma\right\} \\
&\subseteq \left\{\hat\sigma^2 -\sigma^2 > \epsilon^2 ;
  ~\hat\sigma>\sigma\right\} \\ 
&= \left\{\hat\sigma^2 -\sigma^2 > \epsilon^2\right\}.
\end{align*}
Recall that $\hat\sigma^2 = \frac{1}{n}\sum_1^n (X_i - \bar X)^2$. Consider the
unbiased estimator for $\sigma^2$, $S^2 = \frac{n}{n-1}
\hat\sigma^2$, and 
$$\left\{\hat\sigma^2 -\sigma^2 > \epsilon^2\right\} \subseteq \left\{S^2 -\sigma^2 > \epsilon^2\right\}.$$
By Chebyshev's inequality $\P(S^2 -\sigma^2 > \epsilon^2) \leq
{\rm Var}(S^2)/\epsilon^2$. The rest is a tedious but straightforward
calculation, using $\E(X_1^4)<\infty$, to show that ${\rm Var}(S^2)=O(n^{-1})$.
\end{proof}

\subsection*{Proof that Binmedian Has $O(\log n)$ Expected Number of
  Iterations}

Under the same set of assumptions on the data $X_1,\ldots X_n$, we can
show that the \textsf{binmedian} algorithm has an expected number of
$O(\log n)$ iterations before it converges.  We established that 
$$\E(N_1) \leq \frac{n}{B} 2M(\sigma+\epsilon) + C,$$ and by that same
logic it follows
$$\E(N_j|N_{j-1}=m) \leq \frac{m}{B} 2M(\sigma+\epsilon) + C = m -
g(m)$$
where $g(m) = m \frac{2M(\sigma+\epsilon)(B-1)}{B}-C$.
By Theorem 1.3 of \citeasnoun{randalg}, this means that 
\begin{align*}
\E(T) &\leq \int_1^n \frac{dx}{g(x)} \\
&= \int_1^n \frac{dx}{x \frac{2M(\sigma+\epsilon)(B-1)}{B} - C} \\
&= O(\log n). \\
\end{align*}

\section*{Acknowledgements}
We could not have done this work without Rob Tibshirani's positive encouragement and consistently great input.  We also thank Jerry Friedman for many helpful discussions and for his mentorship.  Finally, we thank the Cytobank group --- especially William Lu --- for discussing ideas which gave rise to the whole project.  \textsf{Binapprox} is implemented in the current version of the Cytobank software package for flow cytometry data analysis (www.cytobank.org).

\bibliographystyle{agsm}
\bibliography{tibs}

\end{document}